\documentclass[a4paper,11pt]{article}
\usepackage{pos}
\usepackage{graphicx}
\usepackage{amsmath}
\usepackage{amsthm}
\usepackage{amssymb}
\usepackage{lineno}
\usepackage{caption}
\usepackage{subcaption}
\usepackage{multirow}
\usepackage{multicol}
\usepackage{tabularx}
\usepackage{xcolor}
\usepackage{physics}
\usepackage{wrapfig}
\usepackage{lineno}
\usepackage{multicol}
\usepackage{subfiles}

\usepackage{color}
\usepackage{hyperref} 
\hypersetup{colorlinks=true,allcolors=blue}

\title{Detection methods for the Cherenkov Telescope Array at very-short exposure times}
\ShortTitle{Short-term detection with CTA}

\author*[a]{Ambra Di Piano}\author[a]{Andrea Bulgarelli}\author[a]{Valentina Fioretti}
\author[a]{Leonardo Baroncelli}\author[a]{Nicolò Parmiggiani}\author[b,c]{Francesco Longo}\author[d]{Antonio Stamerra}\author[e,f]{Alicia López-Oramas}\author[a,g]{Giulia Stratta}\author[a]{Giovanni De Cesare}

\affiliation[a]{INAF - Osservatorio di Astrofisica e Scienza dello Spazio di Bologna, Italy}
\affiliation[b]{Università degli Studi di Trieste, Italy}
\affiliation[c]{INFN Sezione di Trieste, Italy}
\affiliation[d]{INAF - Osservatorio Astronomico di Roma, Italy}
\affiliation[e]{Instituto de Astrofisica de Canarias, La Laguna, Spain}
\affiliation[f]{Departamento de Astrofisica, Universidad de La Laguna, Spain}
\affiliation[g]{INFN, Sezione di Firenze, Italy}

\forColl{CTA Consortium} 

\emailAdd{ambra.dipiano@inaf.it}

\abstract{
The Cherenkov Telescope Array (CTA) will be the next generation ground-based observatory for very-high-energy (VHE) gamma-ray astronomy, with the deployment of tens of highly sensitive and fast-reacting Cherenkov telescopes. It will cover a wide energy range (20 GeV - 300 TeV) with unprecedented sensitivity. To maximize the scientific return, the observatory will be provided with an online software system that will perform the first analysis of scientific data in real-time. This study investigates the precision and accuracy of available science tools and analysis techniques for the short-term detection of gamma-ray sources, in terms of sky localization, detection significance and, if significant detection is achieved, a first estimation of the integral photon flux. The scope is to evaluate the feasibility of the algorithms' implementation in the real-time analysis of CTA. In this contribution we present a general overview of the methods and some of the results for the test case of the short-term detection of a gamma-ray burst afterglow, as the VHE counterpart of a gravitational wave event.
}

\FullConference{37$^{\rm{th}}$ International Cosmic Ray Conference (ICRC 2021)\\
		July 12th -- 23rd, 2021\\
		Online -- Berlin, Germany}

\begin{document}
\maketitle

\section{Introduction}
\label{sec:intro}
The Cherenkov Telescope Array (CTA) will be the next generation of Imaging Atmospheric Cherenkov Telescopes (IACTs) and the largest ground-based gamma-ray detection observatory of the next decade. IACTs operate by observing the Cherenkov radiation induced by the Extensive Air Showers (EAS) produced during the interaction of very-high energy photons with the atmosphere. Data from an array of such telescopes is usually stereoscopically combined to improve the energy and direction reconstruction of the incident gamma-rays.
With dozens of telescopes deployed among two observation sites (in the northern and southern hemispheres), CTA will observe the gamma-ray sky with high energy resolution and unprecedented sensitivity over a broad energy range (20 GeV - 300 TeV). High angular resolution ($\lesssim 0.05^{\circ}$ at $E \geq 1$ TeV) will enable detailed imaging and precise morphology, and a large field of view (up to $\sim8.8^{\circ}$ in diameter) will provide exceptional survey capabilities \cite{ctawhitepaper}. The arrays will couple large effective area with fast slewing capability and unprecedented sensitivity, making CTA a crucial instrument for the future of groud-based gamma-ray astronomy. To maximize the scientific return, the observatory will be provided with an online automated Science Alert Generation (SAG) system \cite{bulgarelli2021} as part of the Array Control and Data Acquisition System \cite{oya2019}.
The SAG will send and receive alerts on transients and variable phenomena (like gamma-ray bursts, active galactic nuclei, gamma-ray binaries, and any serendipitous source) in real-time. The SAG will also provide low-level Cherenkov data reconstruction, data quality monitoring and science monitoring during observations. 
The system is required to search for transient phenomena on multiple timescales (from 10 seconds to 30 minutes) in the field of view, and to issue candidate science alerts with a latency lower than 20 seconds after data acquisition. The sensitivity of the scientific analysis in real-time is nonetheless required to be not worse than half of the sensitivity of the final processing pipelines. Although challenging, these requirements will make the SAG a key system in multi-messenger (MM) and multi-wavelength (MWL) astronomy.
Two science tools are available to the community for the analysis of CTA data, ctools \cite{ctools, ctools2016} and gammapy \cite{gammapy2017, gammapy2019}. Additionally, an aperture photometry tool \cite{tampieri2020} is being developed for the real-time analysis of CTA. To characterize the precision and accuracy of the tools for the detection of candidate sources at the very short exposure (up to 100 s), we inject a simulated observation (comprising gamma-ray photons as well as a diffuse background component due to cosmic ray residuals) and perform a search in the field of view to localize the source. If a candidate is found, we evaluate the significance of the detection and estimate the integrated flux.

\subsection{Aperture photometry}
\label{sec:onoff}
\begin{wrapfigure}{R}{0.5\textwidth}
	\centering
	\captionsetup{width=\linewidth}
	\begin{subfigure}[l]{\linewidth}
		\includegraphics[width=\linewidth]{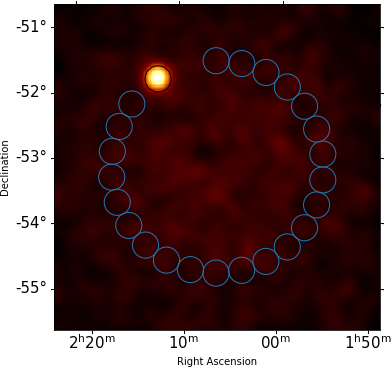}
	\end{subfigure}
	\caption{Example of on source and off source regions defined on a count map with the reflection method, using 0.2$^{\circ}$ region radius and 1.638$^{\circ}$ offset. The areas adjacent to the on regions can be skipped to avoid contamination for the estimation of the background.}
	\label{fig:reflection}
\end{wrapfigure}

The standard on/off analysis for Cherenkov observation includes different approaches to aperture photometry, of which we implement the reflection method.\footnote{We use ctools version 1.7.3, gammapy version 0.18.2 and a photometry tool (version 0.1.0) in development for the SAG.} The on/off technique is based on the extraction of fundamental photometric qualities from a photon list, as the number of photons from a defined region. Conventionally, the aperture (on region) is the region centered on the source itself and is used to count the on-source photons ($N_{\mathrm{on}}$). To estimate the background with the reflection method, one or more off regions with the same characteristic of the aperture (radius and offset from the center of the field of view) are defined and used to count the off-source photons ($N_{\mathrm{off}}$). The photon excess is computed as $N_{\mathrm{{S}}} = N_{\mathrm{on}} - \alpha N_{\mathrm{off}}$, where $\alpha$ is the background scaling factor:

\textcolor{black}{
\begin{equation}
\label{eq:alpha}
\alpha = \frac{A_{\mathrm{on}} \cdot t_{\mathrm{on}} \cdot k_{\mathrm{on}}}{A_{\mathrm{off}} \cdot t_{\mathrm{off}} \cdot k_{\mathrm{off}}},
\end{equation}
}

\noindent where $A$ is the effective area, $t$ the exposure, and $k$ the size of the region. The reflection method allows to define on and off regions in the same observation, reducing eq~(\ref{eq:alpha}) to $\alpha = 1 / N_{\mathrm{reg}}$, with $N_{\mathrm{reg}}$ the number of off regions (an example is shown in figure~\ref{fig:reflection}). The significance is then computed via the analytic Li and Ma \cite{lima} formula:

\textcolor{black}{
\begin{equation}
\label{eq:lima}
S = \sqrt{2}\Big\{N_{\mathrm{on}} \ln{\Big[\frac{1+\alpha}{\alpha}}\Big(\frac{N_{\mathrm{on}}}{N_{\mathrm{on}}+N_{\mathrm{off}}}\Big)\Big] + N_{\mathrm{off}} \ln{\Big[(1+\alpha)\Big(\frac{N_{\mathrm{off}}}{N_{\mathrm{on}}+N_{\mathrm{off}}}\Big)}\Big] \Big\}^{1/2}.
\end{equation}
}

\subsection{Full field-of-view maximum likelihood}
\label{sec:fullfov}
Alternatively to the standard on/off analysis, we implement an unbinned full field-of-view analysis\footnote{We use ctools version 1.7.3; a binned 3d analysis is being developed with the use of gammapy.} for the significance evaluation of the detected candidate. The analysis performs a maximum likelihood fit using the Poisson formula for maximum likelihood estimation (MLE) given the reconstructed direction $\Vec{p}'$, the measured energy $E'$ and the trigger time $t'$.

\textcolor{black}{
\begin{equation}
-\ln{L(M)} = e(M) - \sum_{\mathrm{k}}{\ln{P(\vb*{p}'_{\mathrm{k}}, E'_{\mathrm{k}}, t'_{\mathrm{k}}|M)}},
\label{eq:max-like-poisson}
\end{equation}
}

\noindent where the maximum likelihood function $L(M)$ describes the probability of the collected data during the observation to be drawn from a particular model $M$, $P$ is the probability density conditioned to a given model $M$ at each event $k$, and $e$ represent the total number of predicted events expected to occur given the model $M$.
The source model comprises of two components: a simple power law spectral model and a point-like source spatial model with extension determined by the Point Spread Function (PSF) of the detector. The background rate is provided by the Instrument Response Functions (IRF) as function of off-axis angle and energy. By convolution with the IRF, the maximum likelihood fit adjusts a subset of parameters in order to find the values that best represent the measured data. The detection significance of the source model is described by a Test Statistic (TS) value:

\textcolor{black}{
\begin{equation}
\label{eq:TS}
TS = 2(\ln{L(M_{\mathrm{s}}+M_{\mathrm{b}}}) - \ln{L(M_{\mathrm{b}})}),
\end{equation}
}

\noindent where $\ln{L(M_{\mathrm{s}}+M_{\mathrm{b}})}$ is the log-likelihood value obtained when fitting the source and the background together to the data, and $\ln{L(M_{\mathrm{b}})}$ is the log-likelihood value obtained when fitting only the background model to the data. The number $n$ of degrees of freedom (dof) of the analysis is the number of free parameters in the source model. In this study, the pipeline is run at $n=1$, with the coordinates and spectral index of the candidate's model fixed, and the power law normalization free. For $n=1$ dof, we verified that the relation $\sigma\approx\sqrt{TS}$ holds also for very-short exposure times (down to 1 s).


\section{Application to a BNS merger}
\label{sec:sim}
In this contribution, we focus on the SAG short-term reaction to an external alert. Specifically, the application of a short gamma-ray burst afterglow search as counterpart of a gravitational wave event \cite{patricelli2021}. 
The goal is to verify the agreement between analyses performed with the same techniques implemented by different science tools, and to constrain the accuracy and precision that can be expected at very-short exposure times for an online automated analysis.
We exploited the GW COSMoS catalogue \cite{patricelli2018, gwcosmos}, a public database of simulated BNS mergers providing the GW signals as detected by the network formed with Advanced LIGO and Advanced Virgo \cite{GWnet}. Each GW detection comes with a sky localization probability map, for given distance and inclination of the orbital plane of the BNS.
To simulate the electromagnetic counterpart of a BNS merger, we use the associated afterglow template that provides the high energy emission \cite{nava2014, nava2018} given the GRB energy, redshift and viewing angle. The intrinsic spectral model is a simple power law, with normalization varying throughout the temporal evolution. 
We select a BNS merger with localization uncertainty comparable to the CTA field of view, located at a redshift of 0.097. The electromagnetic counterpart is at 1.638$^{\circ}$ off-axis angle from the peak of the sky localization probability map (R.A. = 31.582 and DEC. = -53.211 degrees) that we set as pointing coordinates. The isotropic energy of the counterpart is $E_{iso}=1.48 \cdot 10^{51}$ erg, with intrinsic spectral shape of a simple power law of photon index -2.1 and normalization varying from 2.45$\cdot10^{-7}$ to 3.1$\cdot10^{-15}$ (ph$\,$cm$^{-2}\,$s$^{-1}\,$GeV$^{-1}$) in its temporal evolution. We add an exponential cut off to the intrinsic source spectral model, to account for the Extra-galactic Background Light absorption as $F^{ebl}(E)=F(E) \cdot e^{-\tau(E)}$, where $\tau(E)$ is the optical depth value from Ref.~\cite{gilmore2012}.

\begin{figure}[ht]
	\centering
	\captionsetup{width=\linewidth}
	\begin{subfigure}[l]{0.32\linewidth}
		\includegraphics[width=\linewidth]{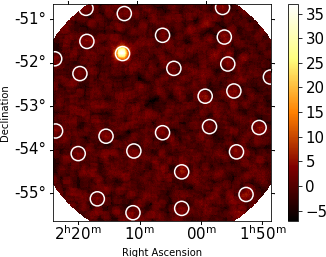}
		\caption{}
	\end{subfigure}
	\begin{subfigure}[c]{0.32\linewidth}
		\includegraphics[width=\linewidth]{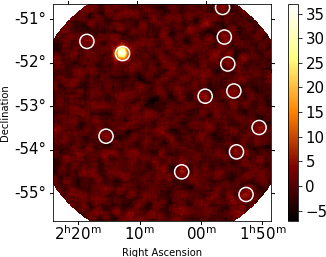}
		\caption{}
	\end{subfigure}
	\begin{subfigure}[r]{0.32\linewidth}
		\includegraphics[width=\linewidth]{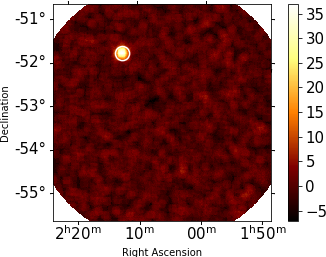}
		\caption{}
	\end{subfigure}
	\caption{Source localization of a simulated gamma-ray burst afterglow ($E_{iso}=1.48 \cdot 10^{51}$ erg at $z=0.097$) using gammapy, in the energy range 40 GeV - 150 TeV using the 30 minutes CTA South IRF at 40$^{\circ}$ of zenith angle. The panels show a TS map with the number and position of hot-spots localized by the peak-search algorithm in a 10 s time window, requiring a significance threshold of (a) 3$\sigma$, (b) 5$\sigma$, and (c) 8$\sigma$.}
	\label{fig:detection}
\end{figure}

\section{Source localization}
\label{sec:localise}
The analysis takes a simulated photon list as input, with given configuration for the energy range, time interval, region of interest, and Instrument Response Function\footnote{\url{https://www.cta-observatory.org/science/cta-performance/}} for the analysis. The sky localization of the candidate source is performed in the field of view of the observation. We assume an extra-galactic scenario, therefore the background is mostly due to cosmic ray induced events that survive the gamma-ray selection criteria during the Cherenkov reconstruction. 
We perform a peak search to localize the coordinates of hot-spots with significance above a given acceptance threshold, selecting the most significant as the candidate source. The algorithm accepts exclusion regions to mask known sources in the field of view.  Figure~\ref{fig:detection} is an example of source localization in a 10 s time window, with different significance acceptance thresholds. At the very short exposure time, the background fluctuation becomes relevant due to the low counting rate and several hot spots are therefore detected. 
We evaluate the localization accuracy as the peak value of a Rayleigh distribution describing the on-sky distance between the detected and true coordinates of the source, whilst the precision is given by the R$_{68}$ containment radius of $10^3$ Monte Carlo realizations of the same source event. While the sigma acceptance threshold has no impact on either accuracy and precision of the localization, parameters such as the pixel size of the sky map required to run the peak-search do. 
\begin{wrapfigure}{r}{0.7\textwidth}
	\centering
	\captionsetup{width=\linewidth}
	\begin{subfigure}[c]{\linewidth}
		\includegraphics[width=\linewidth]{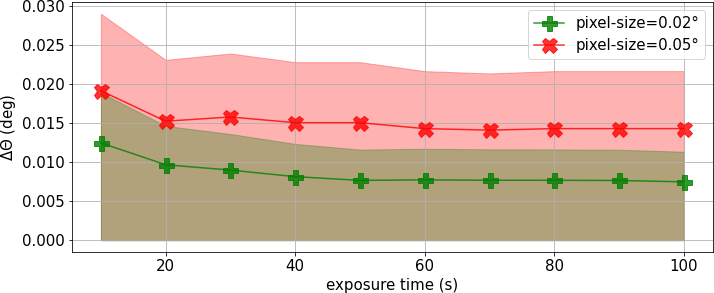}
	\end{subfigure}
	\caption{Comparison between the accuracy and precision of the source localization with pixel size of 0.02$^{\circ}$ (green plus markers) and 0.05$^{\circ}$ (red cross symbols) using ctools, in the energy range 40 GeV - 150 TeV using the 30 minutes CTA South IRF at 40$^{\circ}$ of zenith angle. The panel shows the peak on-sky distance ($\Delta \Theta$) between the true and detected coordinates of the source, with relative R$_{68}$ containment radius (shaded areas).}
	\label{fig:pixsize}
\end{wrapfigure}
In figure~\ref{fig:pixsize} we present an example of the on-sky distance distribution at increasing exposure time (from 10 to 100 s) using a pixel size of 0.02$^{\circ}$ and 0.05$^{\circ}$ with respect to the difference in computational time required to complete the task. A finer spatial binning results in better accuracy and precision by a factor of 2, with little to no impact on the computational time required to complete the task ($\sim0.001$ s).

\begin{figure}[ht]
	\centering
	\captionsetup{width=\linewidth}
	\begin{subfigure}[c]{\linewidth}
		\includegraphics[width=\linewidth]{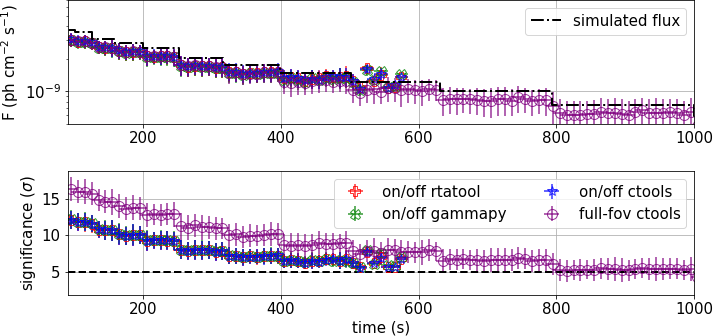}
		\caption{}
	\end{subfigure}
	\caption{Lightcurves of a gamma-ray burst afterglow with 10 s of time window, computed with a full field-of-view maximum likelihood analysis (purple circles) and standard on/off analysis (red crosses, green exes ad blue stars) implemented by different science tools, in the energy range 40 GeV - 150 TeV using the 30 minutes CTA South IRF at 40$^{\circ}$ of zenith angle. The top panel shows the temporal evolution of the flux (the simulated flux is represented by the dashed line), and the bottom panel provides the significance of each detection. The lightcurves last for as long the significance is above 5$\sigma$.}
	\label{fig:tools}
\end{figure}

\section{Significance and flux estimation}
\label{sec:sigmaflux}
In figure~\ref{fig:tools} we present lightcurves and detection significance with 10 s time windows, computed with a maximum  likelihood analysis implemented with ctools, and the on/off reflection analysis implemented with ctools, gammapy and the real-time photometry tool \cite{tampieri2020}.

\begin{wrapfigure}{r}{0.7\textwidth}
	\centering
	\captionsetup{width=\linewidth}
	\begin{subfigure}[c]{\linewidth}
		\includegraphics[width=\linewidth]{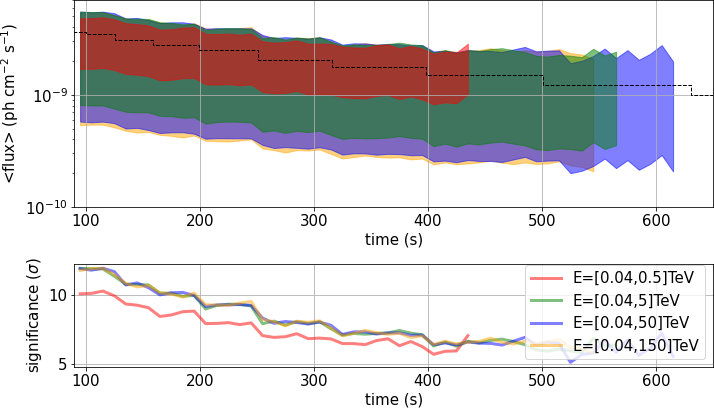}
	\end{subfigure}
	\caption{Impact of the choice of the photon index in the flux estimation with the SAG photometry tool, for given analysis configuration. The lightcurves have 10 s time windows. In the top panel the flux interval (shaded area) is computed with photon indexes between -1.5 and -3 for different energy ranges, compared to the simulated lightcurve (lines) using the 30 minutes CTA South IRF at 40$^{\circ}$ of zenith angle. In the bottom panel the significance is provided. The lightcurves last as long as the significance is above 5$\sigma$.}
	\label{fig:indexes}
\end{wrapfigure}
The choice of the photon index introduces the largest uncertainty in the flux estimation, due to the EBL absorption becoming increasingly relevant at higher energies. For the flux computation a simple power law with photon index -2.4 has been assumed as spectral model for the source, while the simulated model is a cutoff power law with spectral index -2.1. Due to the wide energy range and the EBL absorption not taken into account, the model assumed for reconstruction of the flux introduces a bias. Figure~\ref{fig:indexes} shows the integrated flux intervals computed with photon index between -1.5 and -3 for several energy ranges. The smallest energy range (0.04-0.5 TeV) improves the accuracy of the flux estimate but causes a loss in detection significance due to the reduced counting rate.

\section{Summary}
We have developed an automated pipeline that handles the analysis of CTA data with different techniques and science tools, to investigate their implementation in an online real-time analysis context. We used ctools and gammapy software packages as well as a photometry tool developed for the real-time analysis. Since with CTA we will be able to produce significant observation at very short exposure time, we focus on the characterization of the short-term reaction of the SAG up to 100 s where statistics is limiting. We verify that the same methods implemented in different science tools agrees under the same assumptions. Given a test case, we find that a finer binning (0.02 deg in pixel size) of the sky map produces a factor 2 more accurate and precise localization of the source with respect to a larger binning (0.05 deg) with negligible loss in terms of computational speed ($\sim 0.001$ s). We compare the source detection significance of different techniques (a full field-of-view maximum likelihood and on/off reflection) and the estimation of the integrated photon flux. The full field-of-view analysis technique is more sensitive than the standard on/off analysis, although the two methods have proven to converge when assuming equal assumptions for the background estimation \cite{fioretti2019}. The standard on/off analysis, though, is computationally faster and provides the significance of a detection independently from model assumptions (i.e. the photon index) and fitting procedure. Due to the assumption of a simple power law spectral model, the impact of an arbitrary fixed photon index causes large uncertainties in the flux estimation mostly due to the EBL absorption that becomes increasingly relevant at higher energies. Future studies will investigate either an optimized choice of photon index (i.e., based on the energy range of the observation, knowledge of the source spectral shape and redshift), improvements on model assumptions or higher degrees of freedom analysis.

\section*{Acknowledgements}
\footnotesize{
\noindent This work was conducted in the context of the CTA Consortium, and it made use of the CTA instrument response functions provided by the CTA Consortium and Observatory.\footnote{\url{http://www.cta-observatory.org/science/cta-performance/}} We
gratefully acknowledge financial support from the agencies and organizations listed here: \url{http://www.cta-observatory.org/consortium_acknowledgments/}.

\noindent This research made use of ctools, a community-developed analysis package for Imaging Air Cherenkov Telescope data based on GammaLib, a community-developed toolbox for the scientific analysis of astronomical gamma-ray data; and gammapy,\footnote{\url{https://www.gammapy.org}} a community-developed core Python package for TeV gamma-ray astronomy.
}

\bibliographystyle{unsrt}
\renewcommand{\bibfont}{\footnotesize}
\bibliography{proceeding2021.bib}
\clearpage

\textbf{The Cherenkov Telescope Array Consortium July 2021 Authors}

H.~Abdalla\textsuperscript{1}, H.~Abe\textsuperscript{2},
S.~Abe\textsuperscript{2}, A.~Abusleme\textsuperscript{3},
F.~Acero\textsuperscript{4}, A.~Acharyya\textsuperscript{5}, V.~Acín
Portella\textsuperscript{6}, K.~Ackley\textsuperscript{7},
R.~Adam\textsuperscript{8}, C.~Adams\textsuperscript{9}
% ...

1 : Centre for Space Research, North-West University, Potchefstroom, 2520, South Africa

2 : Institute for Cosmic Ray Research, University of Tokyo, 5-1-5, Kashiwa-no-ha, Kashiwa, Chiba 277-8582, Japan

3 : Pontificia Universidad Católica de Chile, Av. Libertador Bernardo O'Higgins 340, Santiago, Chile

4 : AIM, CEA, CNRS, Université Paris-Saclay, Université Paris Diderot, Sorbonne Paris Cité, CEA Paris-Saclay, IRFU/DAp, Bat 709, Orme des Merisiers, 91191 Gif-sur-Yvette, France

5 : Centre for Advanced Instrumentation, Dept. of Physics, Durham University, South Road, Durham DH1 3LE, United Kingdom

6 : Port d'Informació Científica, Edifici D, Carrer de l'Albareda, 08193 Bellaterrra (Cerdanyola del Vallès), Spain

7 : School of Physics and Astronomy, Monash University, Melbourne, Victoria 3800, Australia

8 : Laboratoire Leprince-Ringuet, École Polytechnique (UMR 7638, CNRS/IN2P3, Institut Polytechnique de Paris), 91128 Palaiseau, France

9 : Department of Physics, Columbia University, 538 West 120th Street, New York, NY 10027, USA

% ...

%
%
%

\end{document}